\documentclass[aps,prl,twocolumn,superscriptaddress]{revtex4}

\usepackage{graphicx}
\usepackage{amsmath}
\usepackage{amssymb}

\begin{document}

\title{Dynamic Super Efimov Effect}
\author{Zhe-Yu Shi}
\email{shizy07@mails.tsinghua.edu.cn}
\affiliation{Institute for Advanced Study, Tsinghua University, Beijing, 100084, China}
\author{Ran Qi}
\email{qiran@ruc.edu.cn}
\affiliation{Department of Physics, Renmin University of China, Beijing 100872, China}
\author{Hui Zhai}
\email{hzhai@tsinghua.edu.cn}
\affiliation{Institute for Advanced Study, Tsinghua University, Beijing, 100084, China}
\author{Zhenhua Yu}
\email{huazhenyu2000@gmail.com}
\affiliation{School of Physics and Astronomy, Sun Yat-Sen University, Zhuhai, 519082, China}
\affiliation{Institute for Advanced Study, Tsinghua University, Beijing, 100084, China}

\date{\today }

\begin{abstract}

Super Efimov effect is a recently proposed three-body effect characterized by a double-exponential scaling, which has not been observed experimentally yet. Here, we present the general dynamic equations determining the cloud size of a scale invariant quantum gas in a time dependent harmonic trap. We show that a double-log periodicity as the hallmark of the super Efimov effect emerges when the trap frequency is decreased with a specially designed time-dependence. We also demonstrate that this dynamic super Efimov effect can be realized with realistic choices of parameters in current experiments.

\end{abstract}

\maketitle

In 1970, V.~Efimov proposed a remarkable phenomenon which is now known as ``the Efimov effect" \cite{Efimov}. He showed that, for three identical bosons in the three dimensions, when the pairwise interaction between the bosons is short-ranged and in the vicinity of an $s$-wave scattering resonance, there are an infinite number of three-body bound states, whose binding energies $E_n$ obey a discrete geometric scaling with 
\begin{equation}
E_n=E_0 e^{-2\pi n/\tilde s_0},
\end{equation}
where $\tilde s_0\simeq 1.006$ is a \emph{universal} constant \cite{Efimov, review}. The Efimov effect attracts great interests both from the nuclear physics and the cold atom physics communities. In 2005, the first evidence of the Efimov effect was observed in a ultracold Cs gas \cite{Grimm}, and later on more details of the effect were revealed in a series of experiments \cite{Efi_exp1, Efi_exp2, Efi_exp3}. Recently, the Efimov effect has also been detected in the nuclear physics systems \cite{nuclear}. What lies at the heart of the Efimov effect is the geometric scaling behavior. However, if $\tilde s_0$ is small, the exponential dependence makes the geometric scaling difficult to observe, since the next bound state would be very shallow in energy and very large in size. Fortunately, the $\tilde s_0$ parameter can be tuned by the atomic mass ratio in an unequal mass atomic mixture, and it becomes larger for larger mass ratio. Therefore, recently the geometric scaling has been observed in the Li and Cs mixture \cite{geometric1, geometric2}.

In 2013, a profound new theoretical development is put forward by Nishida, Moroz and Son \cite{son}. They showed that, for three identical fermions confined in two dimensions, when the pairwise interaction is at the vicinity of a $p$-wave resonance, the fermions can also form an infinite number of three-body bound states \cite{son}, called super Efimov states, since the binding energies follow a fascinating double exponential scaling
\begin{equation}
E_n=E_0\exp(-2e^{\pi n/s_0+\theta}),
\end{equation}
where $s_0=4/3$ and $\theta$ is determined by the short-range potential.  
A mathematical proof confirmed the existence of such states \cite{Gridnev}. Although for the super Efimov states $s_0$ can also be tuned by mass ratio \cite{Nishida}, the double exponential scaling makes its experimental observation unprecedentedly challenging. A subsequent hyper-spherical coordinate approach revealed that the super Efimov effect can be understood in terms of a reduced one-dimensional Schr\"odinger equation with an effective potential \cite{zhenhua, zinner}
\begin{equation}
V_{\rm SE}(\rho)\equiv-\frac{1}{4\rho^2}-\frac{s_0^2+1/4}{\rho^2\log^2\left(\rho/\rho^*\right)},\label{vse}
\end{equation} 
where the hyper-radius $\rho$ quantifies the linear size of the super Efimov bound states, and $\rho^*$ is another parameter. From the effective potential Eq.~(\ref{vse}), it becomes clear that the double exponential scaling of the binding energies corresponds to a double-log periodicity of the bound state wave-functions in space \cite{zinner, zhenhua}. 

In a recent separate development, a dynamic Efimov effect for a scale invariant many-body quantum gas has been proposed \cite{us}, in which we considered the situation that the trap frequency $\omega(t)$ decreases as $1/(\sqrt{\lambda}t)$ with time $t$. We have shown that the expanding cloud size as a function of time bears the same geometric scaling behavior as the three-body Efimov effect. This alikeness is due to a profound connection we discovered between the expansion dynamics of a scale invariant quantum gas and a one-dimensional Schr\"odinger equation \cite{us}. This Efimovian expansion has recently been observed experimentally \cite{us}. 

In this paper, we show that the hallmark of the super Efimov effect, the double-log periodicity, emerges in the cloud size of the expanding scale invariant gas when the trap frequency $\omega(t)$ is specifically engineered to vary with time $t$ as
\begin{equation}
\omega_{\rm SE}(t)\equiv\sqrt{\frac{1}{4t^2}+\frac{1}{\lambda t^2\log^2\left(t/t^*\right)}} \label{omegat},
\end{equation}
where $t^*$ and $\lambda$ ($<4$) are tunable parameters. 
Our predicted dynamic super Efimov effect applies to all scale invariant quantum gases, including i) non-interacting bosons or fermions in any dimension, ii) bosonic or fermionic Tonks gases in one dimension, and iii) unitary Fermi gases in three dimensions. Weakly interacting bosons or fermions in two dimensions are approximately scale invariant systems when the quantum anomaly is weak enough to be ignored. Therefore our proposal for detecting the dynamic super Efimov shall be of interest to many cold atom laboratories. Other synthetic systems that can realize a time dependent harmonic trap can also demonstrate this effect. 
Realizing the dynamic super Efimov effect will represent the first, as far as we know, dynamic process in a quantum system with intriguing double-exponential scaling symmetry

\emph{General Formalism.} For a scale invariance gas of $N$ particles in a spherical symmetric harmonic trap with frequency $\omega(t)$, the cloud size quantified by the mean radius square of the gas particles, $\mathcal R^2(t)=\langle\sum_{i=1}^N \mathbf r_i^2\rangle$, satisfies the dynamic equation
\begin{align}
\frac{d^3}{dt^3}\mathcal R^2(t)+4\omega^2(t)\frac{d}{dt}\mathcal R^2(t)+2\frac{d\omega^2(t)}{dt}\mathcal R^2(t)=0, \label{dynamic}
\end{align}
which can be derived by calculating directly the equation of motion for $\mathcal R^2(t)$ \cite{us, gao}. The absence of interactions in Eq.~(\ref{dynamic}) is due to the emergent Schr\"odinger symmetry in the scale invariant systems \cite{Pitaevskii}.
The solution to Eq.~(\ref{dynamic}) can be expressed as 
\begin{equation}
\mathcal{R}^2(t)=C_1 f^2_1+C_2 f_1f_2+C_3 f^2_2,
\end{equation}
where $f^2_1$, $f_1f_2$ and $f^2_2$ are three linearly independent functions, and $f_{a}$ ($a=1,2$) are two linearly independent solutions of the second order differential equation 
\begin{equation}
\frac{d^2}{dt^2}f+\omega^2(t)f=0. \label{df}
\end{equation}
Here $C_a$ ($a=1,2,3$) are constants to be determined by the initial conditions. 

A hallmark of the super Efimov effect is that 
the zero-energy wave-function $\psi$, satifying $-d^2\psi/d^2\rho+V_{\rm SE}(\rho)\psi=0$, is double-log periodic in $\rho$ \cite{zhenhua, zinner}. 
Comparing Eq.~(\ref{df}) with Eq.~(\ref{vse}), we can view time $t$ as the hyper-radius $\rho$ and $f(t)$ as the real wave-function $\psi(\rho)$; Eq.~(\ref{df}) becomes a zero-energy Schr\"odinger equation along the radial direction with potential $-\omega^2(\rho)$. The initial conditions of the cloud size at $t=t_0$ ($t_0$ is the time at which the expansion starts) are transcribed to the boundary conditions at small radius $\rho$.

\begin{figure}[t]
	\includegraphics[width=3in]
	{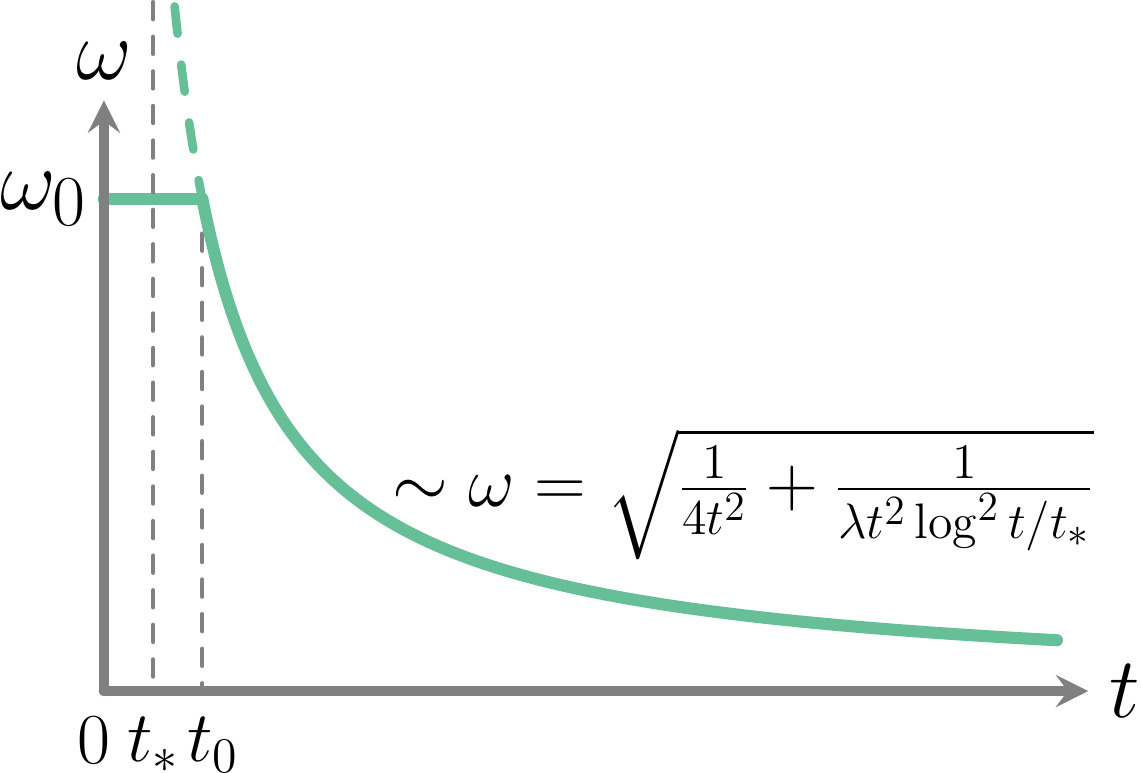}
	\caption{A schematic plot of the time varying trapping frequency $\omega(t)$ for observing the dynamic super Efimov effect.}\label{schematic}
\end{figure}

Therefore, to realize the dynamic super Efimov effect, as shown in Fig.~\ref{schematic}, we consider the situation that the harmonic trap frequency $\omega(t)$ is held at a constant value $\omega_0$ for $t<t_0$, and decreases as Eq.~(\ref{omegat}) for $t>t_0$. The parameters $\omega_0$ and $t_0$ are correlated via $\omega_0=\omega_{\rm SE}(t_0)$. We take the parameter $t^*$ smaller than $t_0$. When $\lambda<4$, the solutions $f(t)$ to Eq.~(\ref{df}), and consequently, $\mathcal{R}^2$, exhibit exactly the same double-log periodicity behavior as predicted for the super Efimov effect \cite{zhenhua, zinner}, which is given by 
\begin{eqnarray}
\mathcal{R}^2=At\log\left(\frac{t}{t_*}\right)\bigg{\{}1+B\cos\bigg{[}s_0\log\left(\log\left(\frac{t}{t_*}\right)\right)+\varphi\bigg{]}\bigg{\}}, \label{super_Efimov}
\end{eqnarray}
where $s_0=2\sqrt{1/\lambda-1/4}$ is tunable by choosing $\lambda$, and $A$, $B$ and $\varphi$ are constants determined by the initial conditions. Since we expand the gas cloud starting from its equilibrium state at $t=t_0$, the initial conditions are
$\mathcal{R}^2{|}_{t=t_0}=\mathcal{R}_0^2$ and $\frac{d}{dt}\mathcal{R}^2{|}_{t=t_0}=\frac{d^2}{dt^2}\mathcal{R}^2{|}_{t=t_0}=0$,
where $\mathcal{R}_0^2$ is the cloud size of the atomic gas at equilibrium when $t<t_0$. We can see from Eq.~(\ref{super_Efimov}) that, on top of a monotonically increasing function $t\log(t/t^*)$, the cloud size does exhibit a double-log periodic oscillation in the time domain. While for $\lambda>4$, the solution becomes
\begin{eqnarray}
\mathcal{R}^2=At\log\frac{t}{t_*}\bigg{[}1+B\bigg{(}\log\frac{t}{t_*}\bigg{)}^\gamma+C\bigg{(}\log\frac{t}{t_*}\bigg{)}^{-\gamma}\bigg{]}, \label{super Efimovian}
\end{eqnarray}
where $\gamma=2\sqrt{1/4-1/\lambda}$; the oscillation disappears.

\emph{Beyond the Super Efimov Effect.}
Based on the same general formalism (\ref{dynamic}-\ref{df}), previously we have successfully realized the dynamic Efimov effect by implementing $\omega(t)=\omega_{\rm E}(t)\equiv1/(\sqrt{\lambda}t)$ with $\lambda<4$ to match the effective hyper-spherical potential $V_{\rm E}(\rho)\equiv-(\tilde s^2_0+1/4)/\rho^2$ \cite{us}, which is known to be responsible for the three-body Efimov effect \cite{Efimov, review}. Equation (\ref{df}) indicates an intriguing underlying connection between the super Efimov effect and the Efimov effect, and suggests interesting effects beyond.

Let us start with the equation 
\begin{align}
\frac{d^2}{dt_1^2} g_1(t_1)+\omega_1^2(t_1)g_1(t_1)=0
\end{align}
with $\omega_1(t_1)=\omega_{\rm E}(t_1)$, which is known to give rise to the dynamic Efimov effect \cite{us}. We introduce the recursive transformation
\begin{align}
t_{n+1}=&e^{t_n},\\
g_{n+1}=&e^{t_n/2}g_n.
\end{align}
We find that $g_n(t_n)$  for $n>1$ generally satisfies 
\begin{align} 
\frac{d^2}{dt_n^2} g_n(t_n)+\omega_n^2(t_n)g_n(t_n)=0
\end{align}
with $\omega_{n+1}^2(t_{n+1})=[\omega_{n}^2(\log t_{n+1})+1/4]/t_{n+1}^2$. For instance, $\omega_2(t)$ is essentially $\omega_{\rm SE}(t)$. In this sense, the dynamic Efimov effect can ``generate" the dynamic super Efimov effect; while $g_1(t_1)$ has a single-log periodicity in $t_1$ \cite{Efimov, review}, $g_2(t_2)$ shows a double-log periodicity in $t_2$. In general, by designing the dynamic variation of the harmonic trapping frequency $\omega(t)$ as $\omega_n(t)$, one can in principle observe a $n$th order log periodicity in the cloud size of a scale invariant gas. While it is relatively easy to engineer $\omega_n(t)$ in the dynamic expansion, the possibility of finding a corresponding effective potential emerging in multi-particle interacting systems is not clear. 

\begin{figure}[t]
	\includegraphics[width=3in]
	{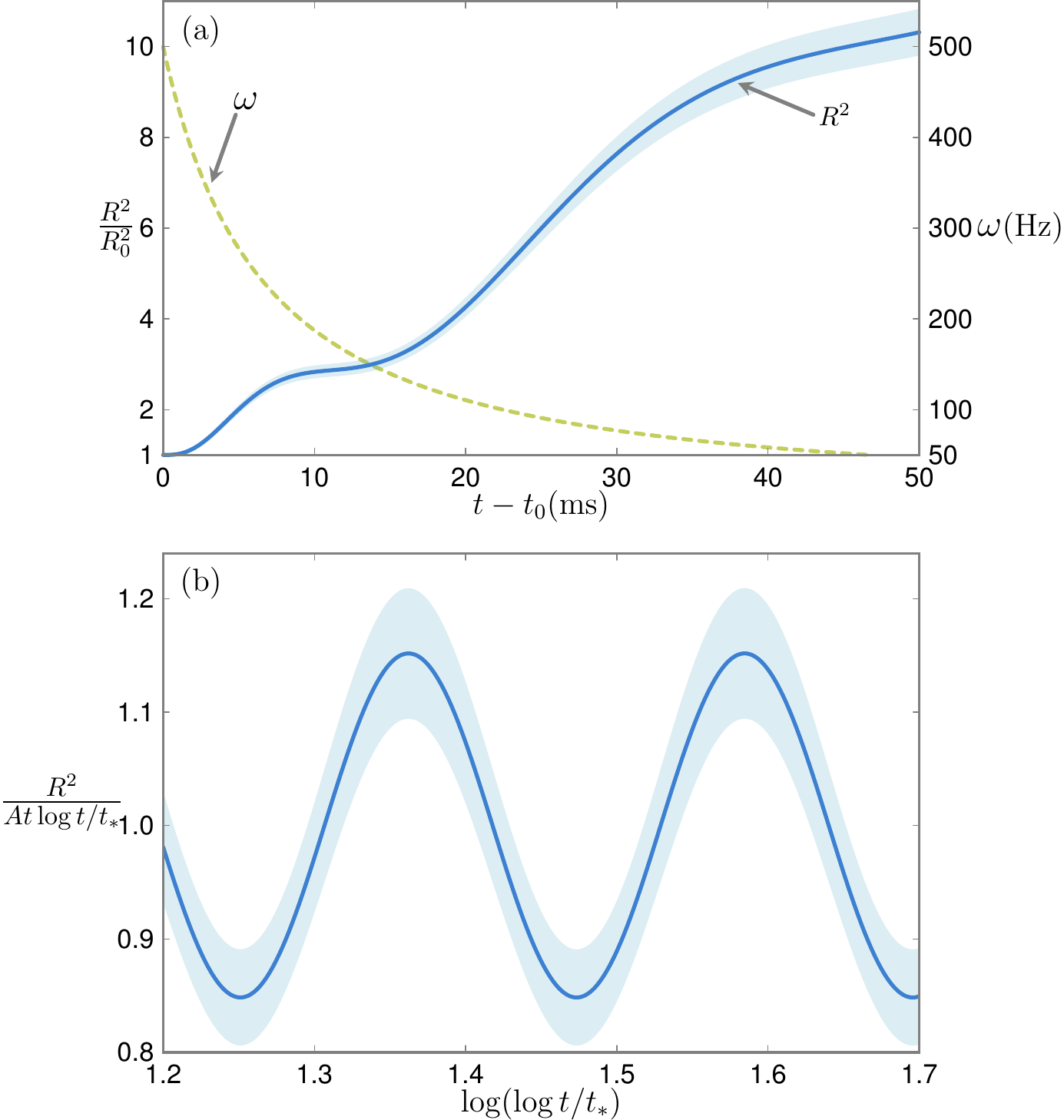}
	\caption{(a) The mean square of the cloud size $\mathcal{R}^2(t)/\mathcal{R}^2_0$, where the dashed line denotes the trap frequency $\omega(t)$ decreases from $500$Hz to $50$Hz. (b) $\mathcal{R}^2(t)/(At\log t/t^*)$ as a function of $\log\log(t/t^*)$. Here we have set $\lambda=0.005$, $t^*=0.3$ms. The shaded area corresponds to an error bar of $5\%$ of the cloud size. }\label{loglogoscillation}
\end{figure}

\textit{Experimental Implementation.} In Eq.~(\ref{omegat}), as $\lambda$ increases, the trap frequency decreases faster with $t$; accordingly, from Eq.~(\ref{super_Efimov}), both the amplitude and the period of the double-log oscillation become larger. In practice, there shall be an upper limit for the initial trap frequency $\omega_0=\omega(t=t_0)$, say, limited by the laser power; there shall also be a lower limit for the final trap frequency $\omega_\text{f}=\omega(t=t_\text{f})$, below which it is hard to control and calibrate the trap frequency against other technical noises. Given these two limits, we should optimize the parameters $\lambda$ and $t_*$, such that i) between $t_\text{f}$ and $t_0$, there are at least 2-3 oscillation periods, i.e., $\mathcal R^2(t)/t\log(t/t^*)$ shows at least three minima/maxima, to identify the double-log periodicity; and ii) the amplitude of the oscillation is large enough that the oscillation is visible despite of the presence of experimental uncertainty in measuring the cloud size. 

In Fig.~\ref{loglogoscillation} we show that when $\omega(t)$ decreases from $500$Hz to $50$Hz within about $50$ms, the mean square of the cloud size $\mathcal{R}^2(t)$ expands about ten times. When we plot $\mathcal{R}^2(t)/(A t\log t/t^*)$, it exhibits a periodic oscillation in term of $\log(\log t/t^*)$. The oscillation completes $>2$ periods within $50$ms and one can identify three minima from this oscillation. In Fig.~\ref{loglogoscillation}(b) we put a shaded area to indicate an error bar of $5\%$ of the cloud size; the error bar is significantly smaller than the oscillation amplitude. With this numerical simulation, we are confident that this dynamic super Efimov effect can be experimentally observed.

\textit{Anisotropic Trap Potentials.} Anisotropy of harmonic traps generically exists in cold atom experiments for dimensions higher than one. Among the aforementioned cold atom systems (i) to (iii) for realizing the super Efimov effect, (i) is a non-interacting system and the equation-of-motion along each trap direction is decoupled, and therefore, it only requires the trap frequency along one of the spatial direction behaving as Eq.~(\ref{omegat}), along which the double-log periodicity can be detected; (ii) is a one-dimensional system; and the trap anisotropy is an issue only for system (iii). 

To take into account the effect of the trap anisotropy on the super Efimovian expansion of system (iii), we consider the situation 
\begin{eqnarray}
\omega_\alpha(t)=\sqrt{\frac{1}{4t^2}+\frac{1}{\lambda_\alpha t^2\log^2t/t_*}}, \label{omegab}
\end{eqnarray}
for $t> t_0$, and $\omega_\alpha(t)=\omega_\alpha(t_0^+)$ for $t<t_0$. Here $\alpha=x,y,z$. The many-body wave-function of the gas $\Psi(\{\mathbf{r}_i\},t)$ satisfies the Schr\"{o}dinger equation
\begin{align}
i\partial_t \Psi=\sum_{i} &\left[-\frac{\nabla_i^2}{2}+\frac{1}{2}\omega_x^2(t)x_i^2+\frac{1}{2}\omega_y^2(t)y_i^2+\frac{1}{2}\omega_z^2(t)z_i^2\right]\Psi\nonumber\\
&+\sum_{i\in\uparrow,j\in\downarrow}V(\mathbf r_i-\mathbf r_j)\Psi,\label{psi}
\end{align}
where the coordinates of the $i$th particle is $\mathbf r_i=(x_i,y_i,z_i)$ and $V$ is the inter-particle interaction potential. We have taken the atomic mass $m=1$. Motivated by the underlying connection between the Efimov effect and the super Efimov effect mentioned above, we perform the following scaling transformation for all spatial-time coordinates and a gauge transformation for the many-body wave-function in the domain $t>t_0$
\begin{eqnarray}
t&=&t_*\exp(e^\tau),\nonumber\\
(x_i,y_i,z_i)&=&(t\log t/t_*)^{1/2}(u_i,v_i,w_i),\nonumber\\
\Psi(\{\mathbf{r}_i\},t)&=&(t\log t/t_*)^{-3N/4}e^{i\theta(\{\mathbf{r}_i\},t)}\Phi(\{\mathbf{q}_i\},\tau),\label{transformation}
\end{eqnarray}
with $\mathbf{q}_i=(u_i,v_i,w_i)$ the scaled coordinates, and the phase function
\begin{eqnarray}
\theta(\{\mathbf{r}_i\},t)=\frac{1}{4t}\bigg{(}1+\frac{1}{\log t/t_*}\bigg{)}\sum_{i}\mathbf r_i^2.
\end{eqnarray}
The transformed many-body wave-function $\Phi$ in the domain $\tau>\tau_0\equiv\log(t_0/t^*)$ satisfies
\begin{eqnarray}
i\partial_\tau\Phi&=&\sum_i\left[-\frac{\nabla_i^2}{2}+\frac{1}{2}\tilde{\omega}_x^2u_i^2+\frac{1}{2}\tilde{\omega}_y^2v_i^2+\frac{1}{2}\tilde{\omega}_z^2w_i^2\right]\Phi\nonumber\\
&&+\sum_{i\in\uparrow,j\in\downarrow}V(\mathbf{q}_i-\mathbf q_j)\Phi,\label{time-independent}
\end{eqnarray}
with
\begin{eqnarray}
\tilde{\omega}_\alpha=
\sqrt{1/\lambda_\alpha-1/4} .
\end{eqnarray}
The interaction part remains the same due to the scale invariance of the potential $V$.

Before the expansion ($t<t_0$), $\Psi(\{\mathbf r_i\},t)$ is an eigenstate of Eq.~(\ref{psi}).
To solve $\Phi(\{\mathbf q_i\},\tau)$ from Eq.~(\ref{time-independent}), the initial conditions can be obtained via Eq.~(\ref{transformation}); $\Phi(\{\mathbf q_i\},\tau_0)$ corresponds to a steady density profile of the gas in a harmonic trap with frequencies $\tilde\omega_\alpha'=\sqrt{1/\lambda_\alpha+\log^2(t_0/t_*)/4}$. 
Equation (\ref{time-independent}) shows that the expansion dynamics has been reformulated in terms of the evolution of the steady gas cloud subjected to a sudden quench in the trap frequencies from $\tilde\omega_\alpha'$ to $\tilde\omega_\alpha$ in each direnctions. When the quench is small, i.e., $1/\lambda_\alpha\gg1$, the excitations on top of the steady gas cloud are dominated by the breathing mode. 
For a typical elongated cylindrical symmetric trap, the breathing mode frequency of the unitary Fermi gas along the long direction (say, $z$-direction) is $\sqrt{12/5}\tilde\omega_z$, which gives rise to the same oscillation frequency in $\sum_i\langle \mathbf q_i^2(\tau)\rangle$.
 Thus, by Eq.~(\ref{transformation}), in this case the mean square of the cloud size $\mathcal R^2(t)$ obeys the same formula as Eq.~(\ref{super_Efimov}) except that $s_0=\sqrt{12/5}\sqrt{1/\lambda_z-1/4}$.

It is worth mentioning that although the above mathematical treatment for the super Efimovian expansion of the unitary Fermi gas in an anisotropic trap is to some degrees similar to that for the Efimovian expansion \cite{us}, the difficulty in the experimental implementation could be quite different. In the Efimovian case, the time-dependence of $\omega_\alpha(t)$ follows $1/(\sqrt\lambda_\alpha t)$. Despite of the different rates in lowering trap frequency along different directions, the aspect ratio of the trap remains a time-independent constant, and therefore, for an optical dipole trap, globally lowering the laser intensity would work. However, the super Efimovian expansion requires that $\omega_\alpha(t)$ follows Eq.~(\ref{omegab}); the aspect ratio of the trap keeps changing during the expansion. Fine-tuning of the laser configuration for an optical dipole trap in such a way poses a challenge in its experimental realization. 

\textit{Summary.} We have presented the general expansion equations governing the cloud size of a scale invariant quantum gas in a time dependent harmonic trap. We have applied the general formalism to predict the emergence of the dynamic super Efimov effect in the expansion when the trap frequency is engineered in the prescribed way.  
We have demonstrated that, with practical experimental setup and parameters, this effect can be observed in non-interacting gases and interacting systems of one-dimensional Tonk gases. For a unitary Fermi gas in three-dimensions, as the trap is generically anisotropic, observing this effect is also possible while more careful control of the trapping frequencies is required. Our study show that the intriguing dynamics of scale invariant gases, combined with the-state-of-art manipulation power of cold atoms, provides a unique chance to observe the super Efimov effect for the first time.

\textit{Acknowledgement.} This work is supported by MOST under Grant No.~2016YFA0301600 (HZ), NSFC Grant No.~11325418 (HZ) and NSFC Grant No.~11474179 (ZY), and Tsinghua University Initiative Scientific Research Program (HZ and ZY), the Fundamental Research Funds for the Central Universities (RQ), and the Research Funds of Renmin University of China under Grant No.~15XNLF18 (RQ) and No.~16XNLQ03 (RQ).

\end{document}